\documentclass[12pt,letterpaper]{article}
\usepackage{jheppub}
\usepackage{verbatim}
\usepackage{slashed}

\usepackage{graphicx}
\usepackage{subfigure}
\input{epsf}
\usepackage{epsfig}
\usepackage{epstopdf}
\usepackage{amsthm}

\usepackage{tikz}
\def\({\left(} \def\){\right)}
\def\[{\left[} \def\]{\right]}
\def\al{\alpha} \def\bt{\beta}
\def\del{{\partial}}

\newcommand{\non}{\nonumber \\}

\newcommand{\be}{\begin{equation}}
\newcommand{\ee}{\end{equation}}
\newcommand{\bea}{\begin{eqnarray}}
\newcommand{\eea}{\end{eqnarray}}
\newcommand{\ba}{\begin{eqnarray}}
\newcommand{\ea}{\end{eqnarray}}

\newcommand{\beq}{\begin{equation}}
\newcommand{\eeq}{\end{equation}}
\newcommand{\beqa}{\begin{eqnarray}}
\newcommand{\eeqa}{\end{eqnarray}}
\newcommand{\beqar}{\begin{eqnarray*}}
\newcommand{\eeqar}{\end{eqnarray*}}

\newcommand{\reef}[1]{(\ref{#1})}

\newcommand{\eg}{{\it e.g.,}\ }
\newcommand{\ie}{{\it i.e.,}\ }

\newcommand{\mt}[1]{\textrm{\tiny #1}}

\title{Linear response of entanglement entropy to $T\bar T$ in massive QFTs}

\author[]{Shachar Ashkenazi$^a$,}
\author[]{Soumangsu Chakraborty$^b$,}
\author[]{Zhanyu Ma$^c$,}
\author[]{Tom Shachar$^a$}
\affiliation[a]{The Racah Institute of Physics, The Hebrew University of Jerusalem, \\ Jerusalem 91904, Israel}
\affiliation[b]{Institut de Physique Th\'eorique, Universit\'e Paris-Saclay, CNRS, CEA\\	
Orme des Merisiers, 91191 Gif-sur-Yvette, France}
\affiliation[c]{School of Physics and Astronomy, Tel Aviv University, Tel Aviv 6997801, Israel}

\emailAdd{shachar.ashkenazi@mail.huji.ac.il}
\emailAdd{soumangsuchakraborty@gmail.com}
\emailAdd{zhanyu.ma@mail.huji.ac.il}
\emailAdd{tom.shachar@mail.huji.ac.il}

\abstract{We compute the leading correction to entanglement entropy in $T\bar{T}$ deformed massive QFTs. We show that both for massive scalar and Dirac fermion, the leading order correction to the entanglement entropy of half space comes from the boundary of the entangling surface. For the case of massive scalar, the boundary term is finite while for massive fermion, it diverges logarithmically giving rise to an additional log-square divergence in the entanglement entropy.
}

\begin{document}
\maketitle
 
 \section{Introduction}
 
 Over the past few years, there have been interesting developments in solvable deformations of two-dimensional quantum field theories (QFT) by a class of irrelevant operators quadratic in the conserved currents \cite{Smirnov:2016lqw,Cavaglia:2016oda}.  Although such deformations involve flowing up the renormalization group (RG), the ultraviolet (UV) divergences are very much under control and the deformed theory turns out to be well-defined. One such solvable irrelevant deformation is the so-called $T\bar{T}$ deformation (where $T$ and $\bar{T}$ are respectively the holomorphic and anti-holomorphic components of the stress tensor) where the seed theory is deformed in such a way that the tangent to the RG trajectory in the space of theories, at all points along the flow, is given by the determinant of the stress-energy tensor of the deformed theory at that point on the RG flow. It was shown in \cite{Smirnov:2016lqw,Cavaglia:2016oda}  that the spectrum of the deformed theory satisfies a flow equation popularly known as the Burgers equation.   Given the spectrum of the undeformed theory, the spectrum of the deformed theory can be uniquely determined in terms of the spectrum of the undeformed theory and the $T\bar{T}$ coupling. The high energy limit of the theory is non-local in the sense that the short-distance behavior is not governed by a fixed point. For one particular sign of the $T\bar{T}$ coupling, the high energy density of states is Hagedorn, while for the other sign of the $T\bar{T}$ coupling, the deformed energies of the excited states (above some threshold) become complex. For an elaborate review of the subject, we will refer to the reader to \cite{Jiang:2019epa} and for generalizations to \cite{Guica:2017lia,Chakraborty:2018vja,Apolo:2018qpq,Chakraborty:2019mdf,LeFloch:2019rut}.
 
The $T\bar{T}$ operator is a composite operator and needs to be defined carefully so that it's free from divergences. It was shown in \cite{Zamolodchikov:2004ce} that in a generic Lorentz invariant local QFT$_2$ on flat space, the $T\bar{T}$ operator defined by point splitting, is unambiguous and non-singular (up to total derivative terms) in the collision limit. Correlation functions in a generic CFT$_2$ deformed by $T\bar{T}$ have been studied in \cite{Cardy:2019qao} to all orders in perturbation theory. In particular, it was shown that the UV divergences can be absorbed into non-local field renormalization to all orders in perturbation theory. Correlation functions of stress tensor operators in a $T\bar{T}$ deformed CFT in a 't Hooft-like limit have been studied in \cite{Aharony:2018vux}, where it was argued that, in this limit, the non-local features of $T\bar{T}$ deformation disappears and the theory can be treated as a local QFT. 
 
 Perturbative renormalization of $T\bar{T}$ deformed QFTs was first studied in  \cite{Rosenhaus:2019utc} where the renormalized Lagrangian was derived from the S-matrix in the case of massive field theories. Renormalization of composite operators in  $T\bar{T}$ deformed QFTs has been studied in \cite{Dey:2020gwm,Dey:2021jyl}. 

There have been a few approaches to understanding entanglement entropy in $T\bar{T}$ deformed CFTs \cite{Donnelly:2018bef,Chakraborty:2018kpr,Asrat:2020uib,Chakraborty:2020udr}. Most of the results available in the literature are holographic. Field theory computation of entanglement entropy of a $T\bar{T}$ deformed CFT is, in general, difficult and not known beyond leading order in perturbation theory. At linear order in the $T\bar{T}$ coupling, the response of entanglement entropy to $T\bar{T}$ vanishes in a CFT \footnote{At linear order, there could be contributions coming from $T$ and $\bar{T}$ contact terms and hence non-universal \cite{Chakraborty:2018kpr,Chakraborty:2020udr}. These contact terms depend on the choice of coordinates in the space of theories and are not given by the CFT data \cite{Kutasov:1988xb}.}, and to the best of our knowledge, the next to leading order correction is not known.  However, the leading order correction to entanglement entropy in a $T\bar{T}$ deformed QFT with a mass scale is non-vanishing. In this paper, we investigate this problem in the case of a free massive scalar field and Dirac fermion. We show that the linear response of entanglement entropy of half space comes from the boundary of the entangling surface. In the case of massive scalar the leading correction has a logarithmic divergence while for massive Dirac fermion we find a log-square divergence along with a log divergence. The presence of the log-square term renders the coefficient of the log term non-universal. 

The paper is organized as follows. In section \ref{section2}, we give a lightning review of the entanglement entropy of half space in a Lorentz invariant QFT. In section \ref{sec3}, we compute the renormalized $T\bar{T}$ operator and the necessary correlation functions needed to compute entanglement entropy. In section \ref{section4}, we evaluate the leading order correction to entanglement entropy and finally in section \ref{section5} we discuss our finding with avenues to future research.

\section{Entanglement entropy of half space}\label{section2}

In this section, we review the results  which are essential for evaluating the entanglement entropy associated with a semi-infinite spatial cut $A=\{(x_1,x_2)\in\mathbb{R}^2: ~ x_1\in (0,\infty)\, , ~x_2=0\}$. It was argued in \cite{Casini:2011kv} that the modular Hamiltonian for this geometry tremendously simplifies. For {\it any} QFT in the vacuum state it can be identified with the generator of rotations around the origin\footnote{In higher dimensions $K$ corresponds to the generator of rotations in the plane orthogonal to the entangling surface. The minus sign on the right hand side ensures $K$ is a positive definite operator for our choice of the stress tensor $T_{\mu\nu}={2\over \sqrt{g}}{\delta I_\mt{E}\over \delta g^{\mu\nu}} $, where $I_\mt{E}$ is the Euclidean action of the theory.}
\be
 K\equiv -\log\rho=-2\pi \int_0^\infty dx_1 x_1 T_{22} + c~,
 \label{modularH}
\ee 
where $\rho=\text{Tr}_{\bar A} |0\rangle\langle0|$ is the vacuum state reduced to $A$, and $c$ is a real normalization constant, such that $\text{Tr}_A ~ \rho=1$. By definition, entanglement entropy is given by
\be
 S_\mt{EE}=-\text{Tr}_A \(\rho \log\, \rho\) = \text{Tr}_A\(\rho K\)=\text{Tr}_A\( e^{-K} K\) = \langle0|K|0\rangle~.
\ee
Representing the vacuum expectation value of the modular Hamiltonian in terms of the Euclidean path integral, one can derive a closed form expression for the variation of entanglement entropy with respect to any coupling $\lambda$ with other  parameters fixed \cite{Rosenhaus:2014woa,Rosenhaus:2014ula,Rosenhaus:2014zza} 
\be
 {\del S_\mt{EE} \over \del \lambda} = - \langle0| K \, {\del I_\mt{E} \over \del\lambda } |0\rangle_c ~,
  \label{1stlaw}
\ee
where the subscript $c$ indicates connected correlator, whereas $I_\mt{E}$ is the full Euclidean action including counterterns which are necessary to generate finite correlation functions. If the normalization condition  $\text{Tr}_A ~ \rho=1$ is not imposed for some reason, the right hand side should be supplemented with an additional term $\langle0| {\del K \over \del\lambda} |0\rangle$. For us this term is absent since the reduced state in the definition of entanglement entropy is normalized.\footnote{$\langle0| {\del K \over \del\lambda} |0\rangle=\text{Tr}_A\( e^{-K} {\del K \over \del\lambda}\)=-{\del \over \del\lambda} \text{Tr} \,\rho =0$.}

%By assumption $\lambda$ in \reef{1stlaw} is a renormalized coupling constant. Hence, insertion of the composite operator ${\del I_\mt{E} \over \del\lambda }=\int d^2 x \,[\mathcal{O}]$ into a correlation function does not generate new type of divergences \cite{?}.  Note however that in general $[\mathcal{O}]$ is different from its bare cousin $\mathcal{O}$ which couples directly to the bare $\lambda_0$ in the action. The difference is encoded in various counterterms depending on the renormalized $\lambda$. We will elaborate the details in the following sections.

Furthermore, since $K$ generates rotations around the origin and can be evaluated on any semi-infinite cut not necessarily with $\theta=0$, it follows that the correlation function of $K$ with any scalar operator is independent of the polar angle $\theta$. It is only a function of the radial distance. In particular,   \reef{1stlaw} can be rewritten as follows
\be
 {\del S_\mt{EE} \over \del \lambda}
 =  (2\pi)^2 \int_0^\infty dr \, r  \int_0^\infty dx_1 x_1  
 \langle0| T_{22}(x_1, x_2=0) 
  \mathcal{O}(r,\theta)~ |0\rangle ~,
  \label{1stEE}
\ee
where $\mathcal{O}(r,\theta)$ is the operator that couples to $\lambda$, whereas $\theta$ can take any value, the final answer is independent of a particular choice. This general formula will be used in section \ref{section4} to evaluate entanglement entropy associated with the so-called $T\bar T$ deformation.

\section{Free massive fields and $T\bar T$}\label{sec3}

In this section, we evaluate  the correlation function $\langle K ~ T\bar T\rangle_c$ for free massive scalar and Dirac fields. which will be used in the following section to calculate the linear response of entanglement entropy to the $T\bar T$ deformation. 

\noindent
Free massive field theories are gaussian, and therefore the stress tensor in the definition of $K$ is quadratic in the fields, whereas $T\bar T$ is quartic in the fields. In particular, to calculate $\langle K ~ T\bar T\rangle_c$ for free fields one has to evaluate the only diagram shown in Fig.\ref{fig:vevTK}. In what follows we consider the massive scalar and Dirac fields separately. 

 \begin{figure}
\begin{center}
\begin{tikzpicture}[scale=1, transform shape]
          \draw (0,0) circle (.8cm);
  \draw (-.8,0) node {$\blacksquare$}; 
    \draw (-.85,-.7) node {$y$}; 
        \draw (-3.58,-.6) node {$x$}; 
  \draw (-2.2,0) ellipse (1.4cm and .8cm);
    \fill[black] (-3.58,0) circle (.18cm);
  \draw (-5.6,0) node {$\langle T_{22}(x)  T\bar T(y)\rangle_c=$};  
\end{tikzpicture}
  \end{center}
   \caption{Diagrammatic representation of $\langle T_{22}(x)  T\bar T(y)\rangle_c$ for free fields. The solid lines represent propagators of either free scalar field or Dirac fermion. Solid dot and square denote insertions of the stress tensor component $T_{22}$ and $T\bar T$ respectively. This correlation function is used to calculate $\langle K ~ T\bar T\rangle$ for free field theories.}
   \label{fig:vevTK}
\end{figure}

\subsection{Massive scalar field theory}

The Euclidean action for a 2D massive scalar field is
\be
 I_\phi=\int d^{2}x\, \( \frac{1}{2} \partial_\mu \phi \partial^\mu \phi +\frac{1}{2} m^2_0 \phi^2\)~.
\ee
The energy-momentum tensor of the theory and the $T\bar T$ operator, denoted by $T\bar T^\phi$, are given by
\bea
 T_{\mu\nu}^\phi&=&\del_\mu \phi \del_\nu\phi -\delta_{\mu\nu} \( \frac{1}{2} \partial_\mu \phi \partial^\mu \phi +\frac{1}{2} m^2_0 \phi^2\)~,
 \label{Tphi}
\\
 T\bar T^\phi(x) &=& \lim_{y\to x} \Big(T_{11}^\phi(x) T_{22}^\phi(y) - T_{12}^\phi(x)T_{12}^\phi(y) \Big)=
 {1\over 4} \Big(m_0^4\phi^4 - \big(\partial_\mu \phi \partial^\mu \phi\big)^2 \Big) ~.
 \label{TT}
\eea
To get the expression for the $T\bar T^\phi$ operator we manipulated \eqref{Tphi} as a classical object. While this approach sounds too na\"{\i}ve, as it may easily overlook contact terms etc., for free fields we expect it to be precise \cite{Rosenhaus:2019utc}. In fact, we could use the definition with points split apart and take the limit of coincident points at the level of the correlation functions. The final result will be exactly the same as one obtained based on \reef{TT}. 

%The deformed action $I_\phi + \delta I$ is defined at some cut off scale $\mu_0$. To ensure perturbative regime we assume the following hierarchy of scales
%\be
% {m_0\over \mu_0} \ll 1 ~, \quad \lambda\mu_0^2 \ll 1 ~.
% \label{hierarchy}
%\ee 
%Furthermore, the derivative with respect to $\lambda$ in \reef{1stlaw} is taken while parameters $m_0^2$ and $\mu_0$ are held fixed. The latter is necessary to obey the definition of the $T\bar T$ deformation.

As shown in Fig.\ref{fig:vevTK}, to calculate the correlator $\langle K ~ T\bar T^\phi\rangle$, we have to contract two fields in \reef{TT}, 
 \be
  T\bar T^\phi\to {1\over 2} \Big(3 m_0^4 \phi^2 \langle \phi^2 \rangle 
  -  \partial_\mu \phi \partial^\mu \phi \langle (\partial \phi)^2 \rangle 
  - 2 \partial_\mu \phi \partial_\nu\phi \langle \partial^\mu \phi \partial^\nu\phi \rangle \Big) ~,
  \label{effTTscalar}
 \ee
where various vevs can be calculated using the standard two point function of the massive scalar field 
\be
  D(x)\equiv\langle \phi (x) \phi (0)\rangle = \int {d^2p\over (2\pi)^2} {e^{i p\cdot x} \over p^2 + m_0^2} ={K_0(m_0 r)\over 2\pi} ~,
  % \underset{m_0r\ll 1}{\longrightarrow} {-1\over 2\pi}\, \log(m_0r) + \ldots~,
  \label{scalar2p}
\ee
\eg
\bea
 \langle \phi^2 \rangle &=&\int^{\mu_0} {d^2p\over (2\pi)^2} {1\over p^2 + m_0^2} = {1\over 2\pi} \log{\mu_0\over m_0} ~,
\label{useful1} \\ \non
 \langle \del_\mu\phi \del_\nu\phi \rangle &=& \int^{\mu_0}  {d^2 p\over (2\pi)^2} {p_\mu p_\nu \over p^2 +m_0^2} 
 ={\delta_{\mu\nu}\over 2} \({\mu_0^2 \over 4\pi} -  m_0^2 \, \langle \phi^2 \rangle \) ~,
 \eea
 where $\mu_0$ is a spherical UV cutoff. Hence,\footnote{We used the following identity to simplify the right hand side of 
 \reef{2pcore}
 \be 2(\del\phi)^2= \del^2\phi^2 - 2\phi \del^2\phi = (\del^2 - 2m^2_0)\phi^2 + 2\phi(-\del^2 + m_0^2)\phi~. \ee 
Note that the last term is proportional to the equations of motion, and therefore can be dropped in the calculation of $\langle K ~ T\bar T\rangle$, because the insertion point of $T\bar T$ does not hit the support of $K$.}
\be
 \langle T^\phi_{22}(x) ~T\bar T^\phi (y)\rangle =  {m_0^2\over 2} \langle\phi^2\rangle
 (\del^2_y + m_0^2) \langle T^\phi_{22}(x) ~ \phi^2(y)\rangle 
 %-{\mu_0^2\over 4\pi} (\del^2_y -2 m_0^2) \langle T^\phi_{22}(x) ~ \phi^2(y)\rangle
 ~,
 \label{2pcore}
 \ee
where we dropped the non-universal power-law divergent terms ($\sim \mu_0^2$). Using \reef{modularH} and \reef{2pcore}, the correlation function of the $T\bar T^\phi$ with the modular Hamiltonian takes the form 
\bea
 \langle K \, ~T\bar T^\phi (r) \rangle_c &=& {m_0^2\over 2} \langle \phi^2 \rangle \({1\over r} {\del\over \del r} r {\del\over \del r} +  m_0^2\) 
 \non
 &\times&{m_0^2\, r^2 \over 2\pi} \Big( K_0^2(m_0r) + {K_0(m_0r) K_1(m_0r)\over m_0r} - K_1^2(m_0r) \Big)~,
 \label{scalarKTT}
\eea
where $r$ is the distance of the insertion point of $T\bar T^\phi$ from the entangling point. Expression in the second line represents $\langle K \, \phi^2 \rangle$ in the massive free field theory. In the next section we use this formula along with \reef{1stlaw} to evaluate the linear response of entanglement entropy to the $T\bar T$ deformation.

\subsection{Massive Dirac field theory}

The Euclidean action for a 2D massive Dirac field has the following form
\be
 I_\psi=\int d^{2}x\,   \bar\psi(\slashed{\del} + m_0)\psi ~.
 \label{I_fermi}
\ee
The energy-momentum tensor of the theory and the $T\bar T$ operator, denoted by $T\bar T^\psi (x)$, take the following form
\bea
 T_{\mu\nu}^\psi&=&{1\over 2} \(\bar\psi\gamma_{(\mu} \del_{\nu)} \psi   - \del_{(\mu} \bar\psi \gamma_{\nu)} \psi\) 
 - \delta_{\mu\nu}  \bar\psi(\slashed{\del} + m_0)\psi ~,
 \label{Tpsi}
\\
 T\bar T^\psi (x) &=&  {1\over 2  } \big(X^\mu_\mu\big)^2 - {1\over 2} X_{\mu\nu} X^{\mu\nu} + m_0 \bar\psi\psi X^\mu_\mu + m_0^2 (\bar\psi\psi)^2 ~,
 \label{TTpsi}
\eea
where the symmetric tensor $X_{\mu\nu}$ is given by\footnote{Parenthesis around a set of indices denotes the symmetrization of a tensor with respect to those indices.}
\be
 X_{\mu\nu} = {1\over 2} \(\bar\psi\gamma_{(\mu} \del_{\nu)} \psi - \del_{(\mu} \bar\psi \gamma_{\nu)} \psi\)~.
\ee
The two point function of the Dirac field can be expressed in terms of its scalar counterpart \reef{scalar2p} 
\be
 \langle \psi(x) \bar\psi(0) \rangle =( - \slashed{\del} + m_0) D(x) ~,
\ee
where $D(x)$ is given by \eqref{scalar2p}.
It can be used to get vevs of various local operators 
\begin{equation}
\begin{aligned} \label{useful2}
 & \langle \psi_\al \bar\psi_\bt \rangle= \delta_{\al\bt} {m_0\over 2\pi} \log{\mu_0\over m_0} ~,\\ 
 & \langle  \psi_\al \del^\nu \bar\psi_\bt \rangle = - \langle \del^\nu \psi_\al \bar\psi_\bt \rangle =  \gamma^\nu_{\al\bt} {m_0^2\over 4\pi} \log{\mu_0\over m_0} ~,\\
 & \langle \del_\mu \psi_\al \del_\nu \bar\psi_\bt \rangle = -\delta_{\mu\nu} \delta_{\alpha\beta} {m_0^3\over 4\pi} \log{\mu_0\over m_0}~,\\
 & \langle \bar\psi \gamma_\mu \del_\nu \psi \rangle = \delta_{\mu\nu} \, {m_0^2\over 2\pi} \log{\mu_0\over m_0} ~,
 \end{aligned}
\end{equation}
where $\mu_0$ is a spherical cut off. Note that we can use the equations of motion, $(\slashed\del + m_0)\psi=0$, while calculating the correlator $ \langle K \, ~T\bar T^\psi \rangle$, because  $T\bar T^\psi$ does not hit the support of the modular Hamiltonian. For instance,
\bea
 && T\bar T^\psi (x)  \quad \sim \quad  - {1\over 2} X_{\mu\nu} X^{\mu\nu} + {1\over 2  } m_0^2 (\bar\psi\psi)^2 ~,
\label{int_simple}
\eea
where $\sim$ means equality up to terms proportional to the equations of motion. 

\noindent
Furthermore, as illustrated in Fig.\ref{fig:vevTK}, to calculate $\langle K ~ T\bar T^\psi\rangle$ we have to contract two out of four fields in the definition of $T\bar T^\psi$ operator, {\eg} contracting two fields in the last term of the above equation, one gets 
\bea
&& {1\over 2  } m_0^2 (\bar\psi\psi)^2 \quad \to \quad 
- {m_0^3\over 2 \pi} \log{\mu_0\over m_0} ~ \bar\psi \psi ~.
\eea
Similarly, using the definition
\bea
 &&X_{\mu\nu} X^{\mu\nu}  =  {1\over 8} \Bigg[ (\bar\psi\gamma_{\nu} \del_{\mu} \psi) (\bar\psi\gamma^{\mu} \del^{\nu} \psi) 
 - (\bar\psi\gamma_{\nu} \del_{\mu} \psi)(\del^\nu \bar\psi\gamma^{\mu} \psi) + (\bar\psi\gamma_{\mu} \del_{\nu} \psi)(\bar\psi\gamma^{\mu} \del^{\nu} \psi)
 \non
 &&  - 2 (\bar\psi\gamma_{\mu} \del_{\nu} \psi)(\del^\nu \bar\psi\gamma^{\mu} \psi) -  (\del_\mu \bar\psi\gamma_{\nu} \psi)  (\bar\psi\gamma^{\mu} \del^{\nu} \psi)
 +(\del_\mu\bar\psi\gamma_{\nu}  \psi)(\del^\nu \bar\psi\gamma^{\mu} \psi) +(\del_\nu\bar\psi\gamma_{\mu}  \psi)(\del^\nu \bar\psi\gamma^{\mu} \psi)\Bigg] ~,
 \nonumber
\eea
and contracting two fields, yields
\be
X_{\mu\nu} X^{\mu\nu}  \to   {m_0^2\over 2}\langle  \bar\psi \psi  \rangle ~ \bar\psi \psi 
+ {1\over 2}  \langle  \bar\psi \psi  \rangle  \del_\nu \bar  \psi \del^\nu \psi 
~ \sim ~  {1\over 4}  \langle  \bar\psi \psi  \rangle \, \del^2\( \bar\psi \psi \) ~,
\ee
where we used the following contraction rules which hold up to terms proportional to equations of motion
\bea
  (\bar\psi\gamma_{\nu} \del_{\mu} \psi) (\bar\psi\gamma^{\mu} \del^{\nu} \psi)\, , ~
  (\del_\mu\bar\psi\gamma_{\nu}  \psi)(\del^\nu \bar\psi\gamma^{\mu} \psi)
 && \to \quad  0 ~,
 \non
   (\bar\psi\gamma_{\mu} \del_{\nu} \psi)(\bar\psi\gamma^{\mu} \del^{\nu} \psi)\, , ~
    (\del_\nu\bar\psi\gamma_{\mu}  \psi)(\del^\nu \bar\psi\gamma^{\mu} \psi)
 && \to \quad m_0^2  \, \langle  \bar\psi \psi  \rangle ~ \bar\psi \psi ~,
  \non
   (\bar\psi\gamma_{\nu} \del_{\mu} \psi)(\del^\nu \bar\psi\gamma^{\mu} \psi) 
 && \to \quad  - \langle  \bar\psi \psi  \rangle \( m_0^2 \bar\psi \psi  +  \del_\nu \bar  \psi \del^\nu \psi \) ~,
 \non
 (\bar\psi\gamma_{\mu} \del_{\nu} \psi)(\del^\nu \bar\psi\gamma^{\mu} \psi) 
  && \to \quad   - \langle  \bar\psi \psi  \rangle ~ 
 \del_\nu \bar  \psi \del^\nu \psi ~.
\eea
Combining, we finally get
\be
 T\bar T^\psi \quad \to \quad 
 - {1\over 2} \langle  \bar\psi \psi  \rangle ~\({1\over 4}\del^2 - m_0^2 \) \bar\psi \psi~.
  \label{TTpsi1}
\ee 
Thus,
 \be
 \langle T^\psi_{22}(x) ~T\bar T^\psi (y)\rangle =   - {1\over 2} \langle  \bar\psi \psi  \rangle ~\({1\over 4}\del^2_y - m_0^2 \)\langle T^\psi_{22}(x) ~  \bar\psi \psi(y)\rangle 
 %-{\mu_0^2\over 4\pi} (\del^2_y -2 m_0^2) \langle T^\phi_{22}(x) ~ \phi^2(y)\rangle
 ~,
 \ee
with
\be
\langle T^\psi_{22}(x) ~\bar\psi \psi(y)\rangle_c =2m_0 \( D(x-y){\del^2 D(x-y) \over \del x^2_2} -  {\del D(x-y) \over \del x_2}{\del D(x-y) \over \del x_2} \)~.
\ee
Using \reef{modularH}, yields
\be
\langle K ~\bar\psi \psi (r)\rangle_c = {m_0\over 4\pi} \({\del^2 \over \del r^2} - 4 m_0^2\) r^2 \Big( K_0^2(m_0r)  - K_1^2(m_0r) \Big) ~,
\ee
and 
\be
 \langle K~T\bar T^\psi (r)\rangle_c =  - {m_0\over 32 \pi} \langle  \bar\psi \psi  \rangle 
 ~\({1\over r} {\del\over \del r} r{\del\over \del r} - 4m_0^2 \)
 \({\del^2 \over \del r^2} - 4 m_0^2\) r^2 \Big( K_0^2(m_0r)  - K_1^2(m_0r) \Big)~,
 \label{DiracKTT}
 \ee
where $r$ is the radial distance from the origin.

\section{Linear response of entanglement entropy to $T\bar T$}\label{section4}
It was shown in \cite{Calabrese:2004eu} that the entangelment entropy in the case of massive $1+1$-dimensional quantum field theory is given by  
\be
 S_\mt{EE}\sim - {c\over 6} \log(m\delta) ~,
 \label{CFT-EE}
\ee
where $c$ is the central charge, $m$ is mass or inverse correlation length, and $\delta$ is the lattice spacing or UV cut off. We expect correction to this formula if the theory is deformed by the $T\bar T$ operator, \ie the coupling $\lambda$ in \reef{1stlaw} is identified with the $T\bar T$ deformation of a quantum field theory. In principle, one can derive the full answer by integrating \reef{1stEE} with $T\bar T$ substituted for $\mathcal{O}$ and using \reef{CFT-EE} as the boundary condition at $\lambda=0$. However, it is difficult to proceed along this guideline in full generality without  additional simplification. For instance, the form of the $T\bar T$ operator is not simple, and far away from a gaussian field theory. Therefore we adopt a perturbative approach. %\ie we  assume the theory is weakly coupled $\lambda m^2\ll1$.

\noindent
The deformed action is defined at some cut off scale $\mu_0$. To ensure the model is weakly coupled and perturbation theory is applicable, we assume the following hierarchy of scales
\be
 {m_0\over \mu_0} \ll 1 ~, \quad \lambda\mu_0^2 \ll 1 ~,
 \label{hierarchy}
\ee 
where $m_0$ are masses of the fields in the undeformed theory. Furthermore, the derivative with respect to $\lambda$ in \reef{1stlaw} is taken while parameters $m_0$, $\mu_0$ etc. are held fixed. The latter is necessary to obey the definition of the $T\bar T$ deformation.

To evaluate the linear response of entanglement entropy to the $T \bar T$ deformation, one should calculate the right hand side of \reef{1stlaw} to zeroth order in $\lambda$. Hence, the stress tensor and $T\bar T$ operator of the undeformed theory should be substituted for $T_{22}$ and $\mathcal{O}$ in \reef{1stlaw}. The corresponding correlation function was evaluated in the previous section and we use it to calculate the linear response of entanglement entropy for scalar and Dirac fields.

\subsection{Scalar field}

To calculate correction to the leading order result \reef{CFT-EE}, one should substitute \reef{scalarKTT} into \reef{1stlaw} and integrate it over the Euclidean 2D plane,
\be
 {\del S_\mt{EE} \over \del \lambda} = -{m_0^2\over 12} \langle \phi^2 \rangle + \pi \, m_0^2 \langle \phi^2 \rangle 
 \lim_{r\to 0} \Big(r {\del\over \del r}\langle K \, \phi^2(r) \rangle \Big) ~,
\ee
where the last term corresponds to the contribution of the total derivative term in \reef{scalarKTT}. Combining with \reef{CFT-EE}, yields
\be
 S_\mt{EE} =  - {1\over 6} \log(m_0/\mu_0) -{\lambda m_0^2\over 12} \langle \phi^2 \rangle + \pi \, \lambda m_0^2 \langle \phi^2 \rangle 
 \lim_{r\to 0} \Big(r {\del\over \del r}\langle K \, \phi^2(r) \rangle \Big) + \ldots ~,
\ee
Using \reef{useful1} and the physical mass of the deformed model, see Appendix \ref{Appx} for definitions, one can rewrite it as follows
\be
S_\mt{EE} =  - {1\over 6} \log(m_\text{ph}/\mu_0) + \pi \, \lambda m_{ph}^2 \langle \phi^2 \rangle 
 \lim_{r\to 0} \Big(r {\del\over \del r}\langle K \, \phi^2(r) \rangle \Big) + \ldots ~,
\ee
The boundary term does not vanish, it is given by
\begin{equation}\label{bdyboson}
 \lim_{r\to 0} \Big(r {\del\over \del r}\langle K \, \phi^2(r) \rangle \Big) =-\frac{1}{2\pi}~.
\end{equation}
Thus the leading order correction to entanglement entropy takes the form
\begin{equation}\label{deltaSboson}
\delta S_{EE}=-\frac{\lambda m_{ph}^2}{2} \langle \phi^2 \rangle=-\frac{\lambda m_{ph}^2}{4\pi}\log\left(\frac{\mu_0}{m_{ph}}\right)~.
\end{equation}

\subsection{Dirac fermion}

Substituting \reef{DiracKTT} into \reef{1stlaw} and integrating over the Euclidean 2D plane, yields
\be
 {\del S_\mt{EE} \over \del \lambda} =   - {m_0\over 12} \langle  \bar\psi \psi  \rangle -  \, {\pi\over 4} \langle \bar\psi \psi \rangle 
 \lim_{r\to 0} \Big(r {\del\over \del r}\langle K \,\bar\psi \psi (r) \rangle \Big) ~,
\ee
where the last term corresponds to the contribution of the total derivative term in \reef{DiracKTT}. Combining with \reef{CFT-EE} and using \reef{useful2} and the definition of the physical mass, see Appendix \ref{Appx}, yields
\be
S_\mt{EE} =  - {1\over 6} \log(m_\text{ph}/\mu_0) -  \, {\pi\over 4} \lambda \langle \bar\psi \psi \rangle 
 \lim_{r\to 0} \Big(r {\del\over \del r}\langle K \,\bar\psi \psi (r) \rangle \Big) + \ldots ~.
\ee
Note that the boundary contribution diverges logarithmically this time
\begin{equation}\label{bdyfermion}
 r {\del\over \del r}\langle K \,\bar\psi \psi (r) \rangle =\frac{m_{ph}}{\pi}\left(1+\gamma_E-\log2+\log m_{ph}  r\right)+O(r\log m_{ph}r)~,
\end{equation}
where $\gamma_E$ is the Eular gamma function. Thus the leading order correction to entanglement entropy takes the form
\begin{equation}\label{deltaSfermion}
\delta S_{EE}=-\frac{\lambda m_{ph}^2}{4\pi}(1+\gamma_E-\log2)\log\left(\frac{\mu_0}{m_{ph}}\right)+\frac{\lambda m_{ph}^2}{4\pi}\left(\log\frac{\mu_0}{m_{ph}}\right)^2~.
\end{equation}
Note that there is an intriguing difference in the leading order correction to entanglement entropy between the deformed free scalar and the Dirac Fermion. The leading order correction in the case of free boson is a log term whose coefficient is universal \eqref{deltaSboson}, whereas in the case of the Dirac fermion, there is a log-square divergence along with a log divergence \eqref{deltaSfermion}. The presence of the log-square terms renders the coefficient of the logarithmically divergent term non-universal.

\section{Discussion}\label{section5}

In this paper, we studied perturbative entanglement entropy in field theories with a mass scale deformed by $T\bar{T}$. It is remarkable that both for massive free boson and Dirac fermion the leading correction to the entanglement entropy comes from the boundary of the entangling surface.  In the case of free massive boson, the boundary contribution \eqref{bdyboson} is finite giving rise to the expected logarithmic divergence in the entanglement entropy \eqref{deltaSboson}. For free massive Dirac fermion on the other hand, the boundary contribution is logarithmically divergent \eqref{bdyfermion} giving rise to an extra log-square divergence in the entanglement entropy \eqref{deltaSfermion}.  

Although the leading order correction to the entanglement entropy, in both cases, are sourced by the boundary terms of the entangling surface, they are not equal when expressed in terms of the physical parameters in the theory and the cutoff scale.  At first glance, this is a bit confusing because one would naively expect an equality of the leading order correction term due to the bosonization duality of two-dimensional QFTs \cite{Calabrese:2004eu,Giveon:2015cgs,Chakraborty:2018kpr}.  However, the bosonization duality has been established in local two-dimensional QFTs by comparing local observables, whereas the  entanglement entropy is a  ``non-local observable'' supported on a region with a boundary. Our finding shows that the bosonization duality holds up to contributions coming from the boundary of the entangling surface.

In an RG flow of a local Lorentz invariant QFT in two dimensions, connecting two local fixed points, the quantity $C\equiv \mu_0\partial_{\mu_0} S_{EE}$ is independent of the regularization \cite{Casini:2006es}. This follows from the fact  that in local QFTs all physical divergent terms are logarithmic.  In the case of $T\bar{T}$ deformed free boson, at leading order, we do see that $C$ is indeed independent of the cutoff scale, but in the case of free Dirac fermions, this is not the case. This is probably due to the fact that  $T\bar{T}$ is an irrelevant operator and its short distance properties are different from local QFTs. It would be interesting to investigate higher order correction to entanglement entropy to see at what order in perturbation theory exotic non-local divergent terms show up in the case of free bosons.

It's well known that the R\'enyi entropy computed for conformal scalars (where the stress tensor is computed by taking derivative with respect to the metric)  and minimally coupled scalars (where the stress tensor is computed by Noether's theorem) differ by contributions from the boundary of the entangling surface \cite{Lee:2014zaa,Casini:2014yca,Herzog:2016bhv}. A  resolution to this discrepancy comes from a careful consideration of the boundary conditions near the entangling surface. In our case the leading correction to the entanglement entropy comes entirely from the boundary of the entangling surface. It would be interesting to analyse the boundary term along the line of \cite{Lee:2014zaa,Casini:2014yca,Herzog:2016bhv} to extract universal features of entanglement in a $T\bar{T}$ deformed massive QFT. This involves various subtleties concerning the boundary terms. We leave it to future research work.

Following our analysis, it would be nice to understand if the boundary term in the entanglement entropy is a universal feature of an arbitrary massive quantum field theory deformed by $T\bar{T}$. 

The holographic dual of a CFT$_2$ deformed by $\lambda T\bar{T}$ is $AdS_3$ gravity with mixed boundary condition \cite{Guica:2019nzm}. The holographic entanglement entropy in such a setup (\ie $AdS_3$ with mixed boundary condition)  has been investigated in \cite{He:2023xnb}. A natural question that arises at this point is how to compute the holographic entanglement entropy in a $ T\bar{T}$ deformed QFT$_2$ with a mass scale. It is well known that for a CFT with mass scale at ({\ie for $\lambda=0$), the holographic entanglement entropy of half space is computed by introducing an IR cutoff in the radial coordinate \cite{Nishioka:2009un}. The radial IR cutoff introduces the mass scale in the problem. It would be interesting to revisit the holographic computation in  \cite{He:2023xnb} with a hard radial IR cutoff and compare with the perturbative results obtained in this paper.

Finally, there  is another integrable irrelevant deformations of two-dimensional CFTs that goes in the name of $J\bar{T}$ deformation \cite{Guica:2017lia,Chakraborty:2018vja}. It would be interesting to calculate the linear response to entanglement due to $J\bar{T}$ deformation. More generally, our analysis can be extended to deformations by a general linear combination of $T\bar{T},J\bar{T}$ and $T\bar{J}$ \cite{Chakraborty:2019mdf,LeFloch:2019rut}.

\section*{Acknowledgements} 

The authors would like to thank
 A. Giveon and M. Smolkin
for helpful discussions and collaboration at early stages of this work.  
The research of S.C. received funding under the Framework Program for Research and
“Horizon 2020” innovation under the Marie Skłodowska-Curie grant agreement n° 945298.
T.S. thanks the Israeli Science Foundation Center of Excellence (grant No. 2289/18) for support of his research.

\appendix

\section{Physical masses}
\label{Appx}

In this appendix we give a brief derivation of physical mass of $T\bar{T}$ deformed massive free boson and Dirac fermion at leading order in perturbation theory. The deformed scalar field theory Lagrangian is given by
\be
 I=I_0+\delta I=\int d^{2}x\, \( \frac{1}{2} \partial_\mu \phi \partial^\mu \phi +\frac{1}{2} m^2_0 \phi^2\) +{\lambda\over 4} \int d^{2}x\,  \Big(m_0^4\phi^4 - \big(\partial_\mu \phi \partial^\mu \phi\big)^2 \Big) ~.
\ee
Its physical mass, $m_{ph}$, up to a linear order in $\lambda$ can be evaluated based on the 1-loop inverse propagator. In momentum space it is given by
\be
D^{-1}(p) = p^2+ m_0^2 + 3 \lambda m_0^4 \langle\phi^2\rangle 
 - 2\lambda \({\mu_0^2 \over 4\pi} -  m_0^2 \, \langle \phi^2 \rangle \) p^2 + \ldots ~.
\ee
Hence,\footnote{Note that the residue of $D(p)$ at $p^2=m_{ph}^2$ is not unity. To get unit residue one needs to do a wave function renormalization. But its contribution will not change the physical mass.}
\be
 m_\text{ph}^2= m_0^2\(1+  { \lambda m_0^2 \over 2\pi} \log{\mu_0\over m_0} +  {\lambda\mu_0^2 \over 2\pi} \)~,
\ee
or equivalently,
\be
 m_0^2= m_\text{ph}^2\(1-  { \lambda m_\text{ph}^2 \over 2\pi} \log{\mu_0\over m_\text{ph}} -  {\lambda\mu_0^2 \over 2\pi} \)~.
 \label{physical mass}
\ee

To get the relation between the physical mass, $m_\text{ph}$, and $m_0$ in the case of Dirac field, one can use the quadratic one-loop effective action\footnote{To liner order in $\lambda$ the $X_{\mu\nu} X^{\mu\nu}$ term does not contribute to the 1-loop effective action, because it is a total derivative.}
\be
 I_\text{1-loop}^\text{eff}=\int d^{2}x\, \[    \bar\psi(\slashed{\del} + m_\text{ph} )\psi  + \ldots \]~,
 \label{I_fermi1}
\ee
where the pole of the two-point function satisfies
\be
 m_\text{ph} = m_0\Big(1 -  {\lambda\, m^2_0\over 2\pi} \log\big({\mu_0\over m_0}\big) \Big)~.
\ee

\newpage

%\bibliography{ref}\bibliographystyle{JHEP}

\providecommand{\href}[2]{#2}\begingroup\raggedright\endgroup

\end{document}